\documentclass[twocolumn,superscriptaddress]{revtex4}
\usepackage{graphicx}
\begin{document}
\title{Kovacs effects in an aging molecular liquid}
\author{Stefano Mossa} 
\affiliation
{
Laboratoire de Physique Th\'eorique des Liquides,
Universit\'e Pierre et Marie Curie , 
4 place Jussieu, Paris 75005, France
}
\author{Francesco Sciortino}
\affiliation
{
Dipartimento di Fisica and INFM Udr and 
Center for Statistical Mechanics and Complexity,
Universit\`a di Roma "La Sapienza", 
Piazzale Aldo Moro 2, I-00185, Roma, Italy
}
\date{\today}
\begin{abstract}
We  study by means of molecular dynamics simulations 
the aging behavior of a molecular model of ortho-terphenyl. 
We find evidence of a a non-monotonic evolution of the volume 
during an isothermal-isobaric equilibration process, a phenomenon 
known in polymeric systems as cross-over (or Kovacs) effect.  
We characterize this phenomenology in terms of landscape properties,
providing evidence that, far from equilibrium, the system explores 
region of the potential energy landscape distinct from the one explored 
in thermal equilibrium. We discuss the relevance of our findings for 
the present understanding of the thermodynamics of the glass state.
\end{abstract}
\maketitle
If two systems {\em in thermodynamical equilibrium} with identical 
chemical composition have the same temperature $T$ and volume $V$ 
we immediately know that they experience the same pressure $P$. 
We know that the two systems will respond in the same way to an 
external perturbation, and that they will be characterized by the 
same structural and dynamical properties. The ability to predict the 
pressure, and the equivalence of structural and dynamical properties, 
derive from thermodynamical principles.

In the case of glasses, systems in out-of-equilibrium conditions, 
the $T$ and $V$ values are not sufficient for  predicting $P$, 
since the state of the system depends on its previous thermal 
and mechanical history.  Different glasses, at the same $T$ and $V$, 
are characterized by different $P$ values. One can ask if the value of
$P$ is sufficient to uniquely define the glass state, i.e., if
two glasses with identical composition having not only the same 
$T$ and $V$ but also the same $P$ are  the same glass.
If this is the case, the two glasses should respond to an external  
perturbation in the same way and should age with a similar dynamics. 

These basic questions are at the hearth of  a thermodynamic understanding 
of the glassy state of matter, and of the possibility of providing a 
theoretical understanding  of out-of-equilibrium systems. Indeed, 
if by specifying $T$, $V$ and also $P$, we uniquely define the glass state 
---its structural and dynamical properties--- it means that it is 
possible to develop an  out-of-equilibrium thermodynamic 
formalism~\cite{davies53,speedy94,cugliandolo97,nieuwenhuizen98,franz00,mossa02b}
where the previous history of the system is encoded in one additional 
parameter. In the interpretation of experimental data, such additional 
parameter is often chosen as a {\it fictive} temperature or pressure, 
in the attempt to associate the glass to a liquid, frozen from a specific 
thermodynamic state.

\begin{figure}[t]
\centering
\includegraphics[width=0.45\textwidth]{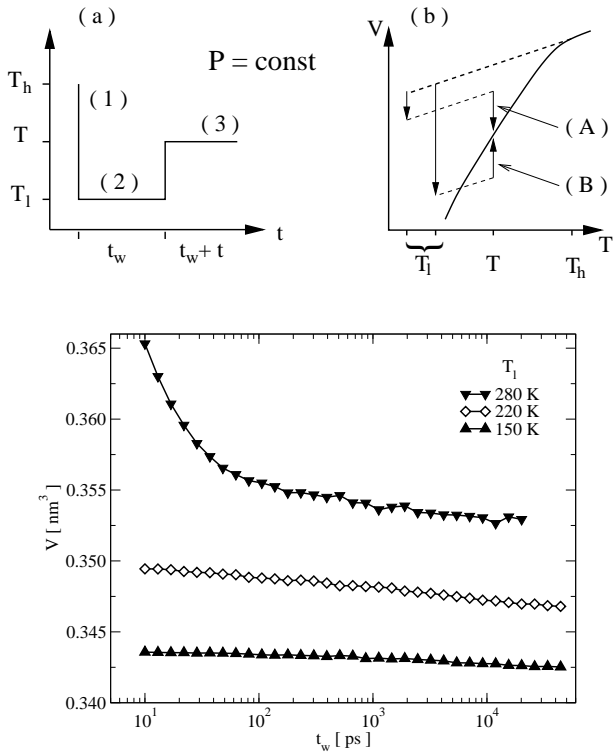}

\vspace{0.7cm}
\includegraphics[width=0.40\textwidth]{fig1b.eps}
\vspace{-0.3cm}
\caption{{\em Top:} {\em (a)} Temperature protocol in Kovacs experiment:
in the numerical simulation the system equilibrated at $T_h$ is 
quenched at several $T_l$ {\em (1)} where it relaxes for a time 
$t_w$ {\em (2)}. It is finally heated at $T$ {\em (3)}.
{\em (b)} Corresponding volume evolution in the cases of low {\em (A)}
and high {\em (B)} $T_l$.
{\em Bottom:} Volume relaxation at different $T_l$. 
The fast relaxation taking place for times shorter than the thermostat 
characteristic time is not reported.}
\label{fig:1}
\end{figure}

Back in the 60th, Kovacs and co-workers designed an experimental 
protocol~\cite{kovacs63,mckenna89,angell00} (Fig.~\ref{fig:1}) 
to generate distinct glasses with different thermal and mechanical 
histories but with the same $T$, $V$ and $P$ values. 
Poly-vinyl acetate was equilibrated at high temperature $T_h$ 
and then quenched at low temperature $T_l$, where it was allowed to  
relax isothermally for a waiting time $t_w$ insufficient to reach 
equilibrium. The material was then re-heated to an intermediate 
temperature $T$, and allowed to relax. The entire experiment was 
performed at constant pressure $P$. The observed dynamics of the 
volume relaxation toward equilibrium---in the last step at constant 
$T$ and $P$--- was striking; the volume crosses over the equilibrium 
value, passes trough a maximum, which depends upon the actual thermal 
history of the system, and then relaxes to the equilibrium value.  
The existence of a maximum clearly indicates that there are states 
with the same $V$ (at the left and at the right of the maximum) which,  
although $T$, $V$ and $P$ are the same, evolve 
differently. Thus, this experiment strongly support the idea that three 
variables are not always sufficient to uniquely predict the state of the 
glass~\cite{nota3}. 

Here we attempt to reproduce numerically the Kovacs experiment,
performing molecular dynamics simulations for a simple molecular model,
to develop an intuition on the differences between states with the
same $T$,$V$ and $P$ and the conditions under which out-of-equilibrium 
thermodynamics may be used to describe glass states. We find that for 
sufficiently deep quenching temperatures, and long aging times, 
following the protocol proposed by Kovacs, it is indeed possible  
to identify two distinct states with the same $T$, $V$ and $P$. 
Thus, for the first time, Kovacs' effects, also know in the literature as
cross-over effects~\cite{scherer86}, are also  observed in 
a molecular liquid model.  Exploiting the possibilities offered by the 
information encoded in  the numerical trajectories, we examine the 
differences between this pair of states, and contrast them with 
corresponding equilibrium liquid configurations.  
We discover that when the system is forced to age following significant 
$T$-jumps, i.e., low $T_l$, it starts to explore regions of the landscape 
which are never  explored in equilibrium.  Under these conditions, 
it is not possible any longer to associate a glass to a "frozen" 
liquid configuration via the introduction of a fictive temperature or 
pressure. This finding limits the range of validity of recent theories 
for out-of-equilibrium systems, based on the possibility of developing a 
thermodynamic formalism for glasses introducing only one 
additional effective parameter in the free 
energy~\cite{nieuwenhuizen98,cugliandolo97,franz00,st01,mossa02b}.

We consider a system of $343$ molecules, interacting via the 
Lewis and Wahnstr\"{o}m  potential~\cite{lewis94} (LW), a model for 
the fragile glass former ortho-terphenyl (OTP). 
The molecule is rigid, composed by three sites located at the vertexes 
of an isosceles triangle.  Sites pertaining to different molecules 
interact by the  Lennard-Jones potential. 
Simulation details are given in Refs.~\cite{mossa02a,lanave02}.  
The system has been studied in isobaric-isothermal conditions;  
the time constants of both the thermostat and the barostat have  
been fixed to $20$ $ps$. In order to accumulate an accurate statistics  
for the analysis below, averages over up to $200$ different starting  
configurations have been performed. The total simulation time is more  
than $6$ $\mu s$.  
\begin{figure}[t]
\centering
\includegraphics[width=0.43\textwidth]{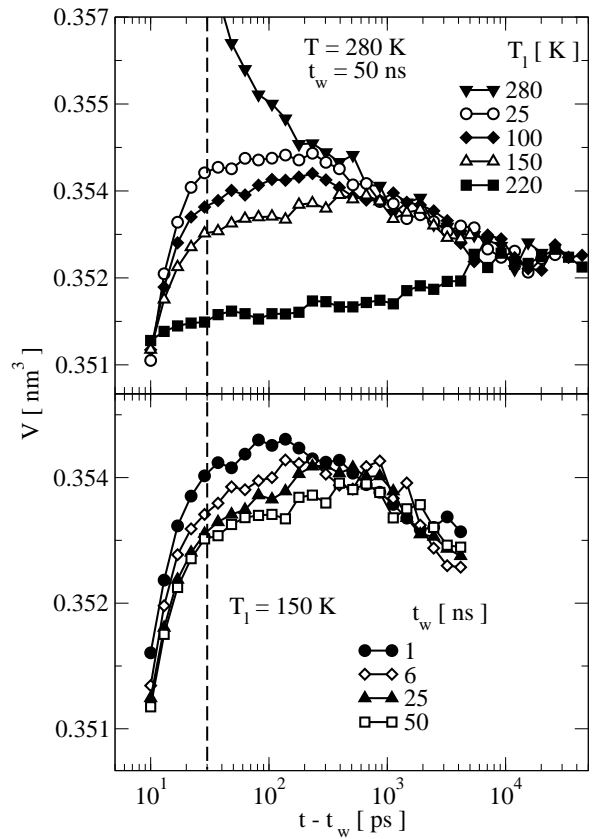}
\vspace{-0.3cm}
\caption{ Volume relaxation showing Kovacs effect in a molecular system.
{\em Top:} A system of  $343$ molecules of the LW 
OTP at constant pressure $P=16$ MPa is equilibrated at $T_{h}=400 K$, 
quenched at several low temperatures $T_{l}$ and leaved to age for 
$t_w = 50$ ns, a time insufficient to reach equilibrium at $T_{l}$.  
The system is then heated at the intermediate temperature $T=280$ K 
and the $V$ relaxation dynamics is recorded (symbols). 
For the case $T_l=T$, the system is directly 
brought from  $T_{h}=400 K$ to $T=280 K$ and hence $t_w=0$. 
The vertical dashed lines indicate the transient time, where $T$ and $P$ 
have not yet equilibrated to the final values.
{\em Bottom:} Volume relaxation at $T=280$ K, at fixed
$T_{l}=150$ K for different $t_w$ values.  
Dashed lines delimitate the transient region, when neither $T$ nor $P$ 
have reached their equilibrium value.
}
\label{fig:kovacs}
\end{figure}

To reproduce the Kovacs experiment, equilibrium configurations 
at $T_{h}=400 \;K$, volume per molecule $V=0.378$ nm$^3$ and $P=16$ MPa 
are isobarically quenched at several low temperatures $T_{l}$, and left 
to age for different $t_w $ values ($t_w$ is never sufficient to 
reach equilibrium at $T_{l}$).  
Each resulting configuration is then isobarically heated 
to $T=280$ K (see Fig.~\ref{fig:1}).  
At $T$ the characteristic structural relaxation time is of the 
order of a few $ns$, allowing us to follow, with the present 
computational resources, the dynamics up to equilibrium.  
The $V$-evolution is recorded along the entire path.   
We also perform an analysis of the properties of the explored 
potential energy landscape (PEL) as a function of time, focusing 
in particular on the energy $e_{IS}$ and pressure $P_{IS}$ of the 
closest local minima configurations (the inherent structures, $IS$), 
as well as the average curvatures of the PEL around the $IS$ configurations.
The $IS$ configuration, which is numerically evaluated performing a 
steepest descent path along the potential energy surface, can be though 
of as the low-$T$ glass generated by instantaneously freezing the liquid.

Fig.~\ref{fig:kovacs}(a) shows the  time evolution of $V$ at 
constant $T=280$ K and $P=16$ MPa for samples which have previously 
aged at different $T_l$ for $t_w=50$ ns. The time evolution 
of $V$  for the case $T_l=T$ (i.e. for a $T$-jump from  $T_h$ to $T$) 
is also reported.
We note that for large $T_l$  (i.e., $T_l=220$ K), $V$ relaxes 
to the equilibrium value from below monotonically. Similarly, for 
$T_l=T$,  $V$ relaxes to the equilibrium value from above, 
again monotonically. At variance, for deep $T_{l}$ values, $V$ does 
not relax  monotonically to the equilibrium value. In analogy 
with Kovacs findings, it goes through a maximum, whose value is higher the 
lower $T_{l}$, before starting to relax toward the equilibrium value. 
After the maximum, the time evolution of $V$ practically coincides 
with case $T_{l}=T$. 

\begin{figure}[t]
\centering
\includegraphics[width=0.48\textwidth]{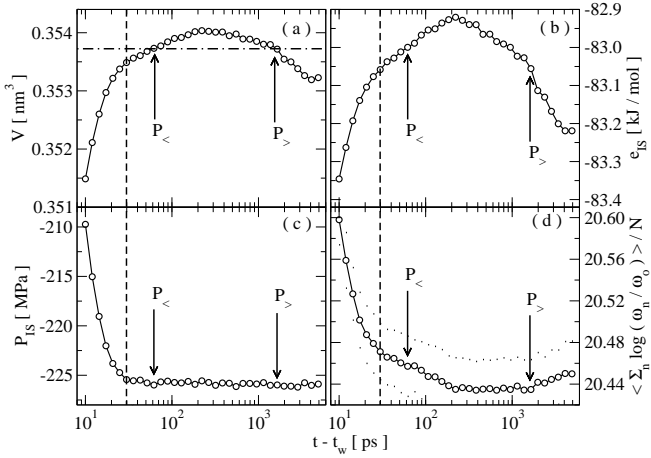}
\vspace{-0.3cm}

\caption{Kovacs effect for $T_{l}=150$ K and $t_w=25$ ns; average over 
$200$ independent realization. The out-of-equilibrium evolution of
{\em (a)} system volume $V$,
{\em (b)} inherent structures $e_{IS}$, 
{\em (c)} inherent structure pressure $P_{IS}$, and 
{\em (d)} shape factor ${\cal S}$
are shown. Error bars are smaller than the symbols. 
Two points (labeled $P_<$ and $P_<$), characterized by the same values 
of $T$, $P$, and $V$  are marked by arrows.}
\label{fig:theory}
\end{figure}

The waiting time dependence of $V$ during the final relaxation at $T$ is shown 
in Fig.~\ref{fig:kovacs}(b) for the case $T_{l}=150$ K.
For all studied $t_w$ values, a maximum is observed, and the
value of the maximum is larger the shorter the waiting time $t_w$. 
All together, Figs.~\ref{fig:kovacs}(a) and (b) confirm that the 
features observed by Kovacs in experiments with polymers are also 
observable in the case of molecular glass forming liquids under 
deep-quench conditions (small $T_l$ values).

Next we study the properties of the region of the PEL explored by 
the system during the the aging process. 
In Fig.~\ref{fig:theory} we show --- for the case $T_l=150$ K for which 
a clear non-monotonic $V$-relaxation is observed ---  the out-of-equilibrium  
evolution of 
{\em (a)} the $V$,
{\em (b)} the average inherent structures energy $e_{IS}$, 
{\em (c)} the average inherent structure pressure $P_{IS}$, and 
{\em (d)} the shape factor  
${\cal S} \equiv \sum_{k=1}^{6N-3}\log(\omega_k/\omega_o)/N$ (here
the  $\omega_k$ are the eigenvalues of the Hessian calculated at the  
inherent structures and $\omega_o$ is  the frequency unit); 
the quantity ${\cal S}$ provides a measure of the 
volume of  the basin of attractions)~\cite{sciortino99}.  
The maximum in the time evolution of $V$ allows us to define
two arbitrary points $P_<$ and $P_>$ (marked in Fig.~\ref{fig:theory}), 
characterized, by construction, by the same $T$, $P$, and $V$ values.  
Although $P_<$ and $P_>$ are indistinguishable from a thermodynamical 
point of view, the subsequent dynamics is completely different in the 
two  cases: after $P_<$ the system expands, while after $P_>$ it 
contracts. From a landscape point of view, $P_<$ and $P_>$ 
differ both in the depth and the shape of the sampled basin. The picture 
that emerges is that, at  $P_<$, the system populates local minima 
which are systematically characterized by energy higher and basins of 
attraction steeper than the ones explored at $P_>$.
\begin{figure}[t]
\centering
\includegraphics[width=0.43\textwidth]{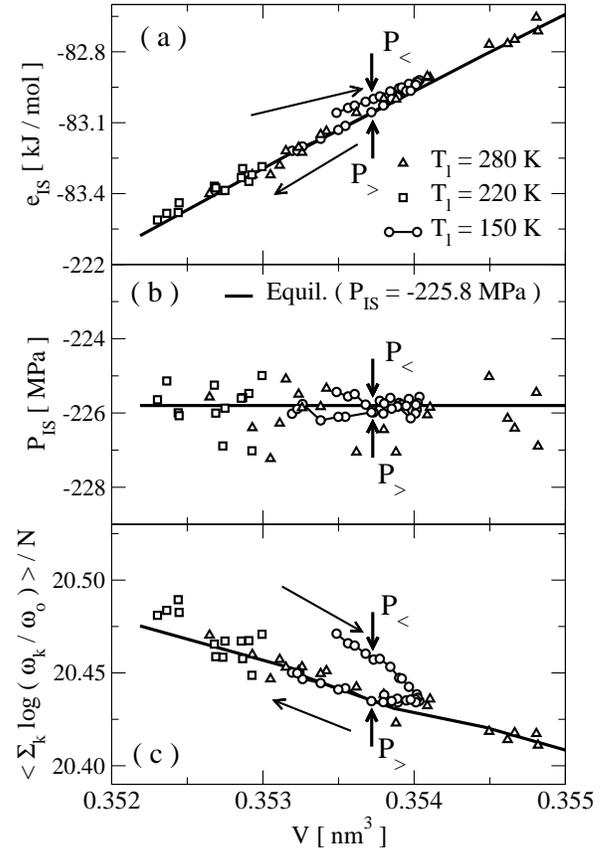}
\vspace{-0.3cm}

\caption{Volume dependence of $e_{IS}${\em (a)} ,$P_{IS}$ {\em (b)}
and ${\cal S}$ {\em (c)}
during the relaxation dynamics at $T=280 K$, for samples which have 
aged at $T_l=150$ K, $220$ K and $280$ K.  
Only data for times longer than $20~ps$
 --- the time associated to the establishment of constant $T$ and $P$ 
conditions --- are shown. Note that $P_{IS}$ is constant.
The solid lines show the equilibrium relation $e_{IS}(V)$, $P_{IS}(V)$ 
and ${\cal S}(V)$ along a constant $P_{IS}=-225.8$ MPa path. 
The vertical arrows indicate $P_<$ and $P_>$ 
(see Fig.~\protect\ref{fig:theory}).} 
\label{fig:finale}
\end{figure}

It is particularly important to note that, after about $20-30$ $ps$, 
corresponding to the time requested to bring  $T$ and $P$ to equilibrium, 
the $P_{IS}$ value stabilizes, while both depth and shape of the explored 
basins continues to change with time. 
This suggests that the vibrational component to the pressure 
(defined as  $P_{vib}=P-P_{IS}$~\cite{mossa02b,lanave03}) 
is rather insensitive to the basin properties, in agreement with similar 
findings from equilibrium studies~\cite{lanave02}. 
The fact that $P_{IS}$ is constant offers an unique possibility to 
estimate how and if the aging dynamics proceeds via a path which is 
usually explored in equilibrium, by comparing properties in equilibrium 
along a constant $P_{IS}$ path and the aging dynamics 
at the same constant $P_{IS}$ value. Indeed, in equilibrium, all landscape 
properties (i.e., $e_{IS}$, $P_{IS}$, ${\cal S}$) are function only of 
$T$ and $V$.   By eliminating $T$ in favor of $P_{IS}$, $e_{IS}$ can 
be expressed as 
a function of $V$ and $P_{IS}$. Therefore, for a fixed value of 
$P_{IS}$, $e_{IS}$ becomes, in equilibrium, an unique function of $V$. 
This equilibrium relation can be compared with the same
relation along the aging path to estimate which part of the
aging dynamics follows the equilibrium properties and
which part is instead different from it. 
A similar comparison can be performed between basin shape ${\cal S}$
and $V$ at constant $P_{IS}$. 

This comparison is shown in Fig.~\ref{fig:finale} for the two cases 
where a monotonic $V$-relaxation is observed (large $T_l$ values, $T_l=220$ K 
and $T_l=280$ K), and for the case $T_l=150 K$ where a maximum in 
the time evolution of $V$ is clearly detected. 
The important information standing out from the comparison is that, 
while equilibration with large $T_l$ values follows equilibrium paths, 
in the other case only the final part of the dynamics follows a path 
which goes via a sequence of states which are explored in equilibrium.  
This confirms that only one $(P_>)$ of the two states with the same 
$T$, $P$, and $V$ can be related, for example via a fictive 
temperature~\cite{expl}, to an equilibrium state, while 
the other one $(P_<)$ has a structure which is never explored in 
equilibrium. Therefore the two states, although having the same 
$T$, $P$, and $V$, are quantitatively different.  
In other words, while at $P_>$ the system samples landscape properties 
which are sampled by the liquid at equilibrium at a higher temperature 
(which can be used as additional parameter to quantify the 
glass properties), at $P_<$ the system explores a region of the landscape 
which is never explored in equilibrium, since the relation between the 
landscape properties characterizing $P_<$ are never encountered in equilibrium. 
Under these conditions, it is not possible to exactly relate the glass 
structure to an equilibrium structure and hence define a fictive 
temperature for the system. 

The present numerical study shows that, only when the change of external
parameters is small, or when the system is close to equilibrium, 
the evolution of the equilibrating system proceeds along a sequence of 
states which are explored in equilibrium. Under these circumstances, 
the location of the aging system can be traced back to an equivalent 
equilibrium state, and a fictive temperature can be defined.  
In this approximation, a thermodynamic description of the aging system 
based on one additional parameter can be provided. When the external perturbation 
is significant, like in hyper-quenching experiments~\cite{angell03}, 
then the aging dynamics propagates the system along a path which 
is never explored in equilibrium.  
In this case it becomes impossible to associate 
the aging system to a corresponding liquid configuration. 
It is a challenge for future studies to find out if a thermodynamic 
description can be recovered decomposing the aging system in a collection 
of sub-states, each of them associated to a different fictive 
temperature---a picture somehow encoded in the phenomenological approaches of 
Tool and co-workers~\cite{tool46} and Kovacs and 
co-workers~\cite{kovacs79}--- or if the glass, produced under extreme 
perturbations, freezes in some highly stressed configuration which can 
never be associated to a liquid state.

We acknowledge several important discussions with W. Kob, E. La Nave and 
G.~Tarjus and support from MIUR PRIN and FIRB. 
We thank G.~Mc Kenna for calling our  attention to the Kovacs experiments.
%
%
%

%
%

\begin{thebibliography}{99}
\bibitem{nota3}
Indeed, the measured kinetics have been rationalized using an 
empirical formalism which postulates the existence of non-exponential 
response functions, and of several retardation mechanisms with different 
time scales~\cite{tool46,kovacs79}.
\bibitem{expl} The fictive temperature could be defined, for example, as 
the temperature at which basins of depth $e_{IS}$ are explored 
in equilibrium, by inverting the thermodynamic relation $e_{IS}(T,V)$
\end{thebibliography}
\end{document}